\documentclass[aps,prl,twocolumn,superscriptaddress,groupedaddress,amsmath,amssymb,nofootinbib,preprintnumbers]{revtex4-2} 

\usepackage{graphicx}
\usepackage{xcolor}
\usepackage{dcolumn}
\usepackage{bm}
\usepackage{subcaption} 
\usepackage{enumitem} 
\usepackage{xspace} 
\usepackage{hyperref}
\usepackage{upgreek}
\hypersetup{
    colorlinks=true,
    linkcolor=cyan,
    filecolor=magenta,      
    urlcolor=cyan,
    citecolor=violet,
}
\usepackage{mathptmx}
\usepackage{cleveref}
\makeatletter
\g@addto@macro\bfseries{\boldmath}
\makeatother

\def\etal{\textit{et~al.}\xspace}
\newcommand{\xmax}{\ensuremath{X_{\mathrm{max}}}\xspace}

\newcommand{\xmaxsigma}{\ensuremath{\sigma(X_{\mathrm{max}})}\xspace}
\newcommand{\xmaxmu}{\ensuremath{\langle X_{\mathrm{max}}\rangle}\xspace}

\newcommand{\gcm}{\ensuremath{\mathrm{g\,cm^{-2}}}\xspace}
\newcommand{\gcmd}{\ensuremath{\mathrm{g\,cm^{-2}\,\mathrm{decade}^{-1}}}\xspace}
\newcommand{\grad}{\ensuremath{^{\circ}}\xspace}

\newcommand{\epos}{EPOS-LHC\xspace}

\begin{document}

\title{Inference of the Mass Composition of Cosmic Rays with energies from $\mathbf{10^{18.5}}$ to $\mathbf{10^{20}}$~eV using the Pierre Auger Observatory and Deep Learning}

\author{
A.~Abdul Halim$^{13}$,
P.~Abreu$^{71}$,
M.~Aglietta$^{53,51}$,
I.~Allekotte$^{1}$,
K.~Almeida Cheminant$^{79,78,69}$,
A.~Almela$^{7,12}$,
R.~Aloisio$^{44,45}$,
J.~Alvarez-Mu\~niz$^{77}$,
J.~Ammerman Yebra$^{77}$,
G.A.~Anastasi$^{57,46}$
% ...
}
\affiliation{}
\collaboration{The Pierre Auger Collaboration}
\email{spokespersons@auger.org}
\author{\phantom{1}}
\affiliation{
\begin{description}[labelsep=0.2em,align=right,labelwidth=0.7em,labelindent=0em,leftmargin=2em,noitemsep,before={\renewcommand\makelabel[1]{##1 }}]
\item[$^{1}$] Centro At\'omico Bariloche and Instituto Balseiro (CNEA-UNCuyo-CONICET), San Carlos de Bariloche, Argentina
\item[$^{2}$] Departamento de F\'\i{}sica and Departamento de Ciencias de la Atm\'osfera y los Oc\'eanos, FCEyN, Universidad de Buenos Aires and CONICET, Buenos Aires, Argentina
\item[$^{3}$] IFLP, Universidad Nacional de La Plata and CONICET, La Plata, Argentina
\item[$^{4}$] Instituto de Astronom\'\i{}a y F\'\i{}sica del Espacio (IAFE, CONICET-UBA), Buenos Aires, Argentina
\item[$^{5}$] Instituto de F\'\i{}sica de Rosario (IFIR) -- CONICET/U.N.R.\ and Facultad de Ciencias Bioqu\'\i{}micas y Farmac\'euticas U.N.R., Rosario, Argentina
\item[$^{6}$] Instituto de Tecnolog\'\i{}as en Detecci\'on y Astropart\'\i{}culas (CNEA, CONICET, UNSAM), and Universidad Tecnol\'ogica Nacional -- Facultad Regional Mendoza (CONICET/CNEA), Mendoza, Argentina
\item[$^{7}$] Instituto de Tecnolog\'\i{}as en Detecci\'on y Astropart\'\i{}culas (CNEA, CONICET, UNSAM), Buenos Aires, Argentina
\item[$^{8}$] International Center of Advanced Studies and Instituto de Ciencias F\'\i{}sicas, ECyT-UNSAM and CONICET, Campus Miguelete -- San Mart\'\i{}n, Buenos Aires, Argentina
\item[$^{9}$] Laboratorio Atm\'osfera -- Departamento de Investigaciones en L\'aseres y sus Aplicaciones -- UNIDEF (CITEDEF-CONICET), Villa Martelli, Argentina
\item[$^{10}$] Observatorio Pierre Auger, Malarg\"ue, Argentina
\item[$^{11}$] Observatorio Pierre Auger and Comisi\'on Nacional de Energ\'\i{}a At\'omica, Malarg\"ue, Argentina
\item[$^{12}$] Universidad Tecnol\'ogica Nacional -- Facultad Regional Buenos Aires, Buenos Aires, Argentina
\item[$^{13}$] University of Adelaide, Adelaide, S.A., Australia
\item[$^{14}$] Universit\'e Libre de Bruxelles (ULB), Brussels, Belgium
\item[$^{15}$] Vrije Universiteit Brussels, Brussels, Belgium
\item[$^{16}$] Centro Brasileiro de Pesquisas Fisicas, Rio de Janeiro, RJ, Brazil
\item[$^{17}$] Centro Federal de Educa\c{c}\~ao Tecnol\'ogica Celso Suckow da Fonseca, Petropolis, Brazil
\item[$^{18}$] Instituto Federal de Educa\c{c}\~ao, Ci\^encia e Tecnologia do Rio de Janeiro (IFRJ), RJ, Brazil
\item[$^{19}$] Universidade de S\~ao Paulo, Escola de Engenharia de Lorena, Lorena, SP, Brazil
\item[$^{20}$] Universidade de S\~ao Paulo, Instituto de F\'\i{}sica de S\~ao Carlos, S\~ao Carlos, SP, Brazil
\item[$^{21}$] Universidade de S\~ao Paulo, Instituto de F\'\i{}sica, S\~ao Paulo, SP, Brazil
\item[$^{22}$] Universidade Estadual de Campinas (UNICAMP), IFGW, Campinas, SP, Brazil
\item[$^{23}$] Universidade Estadual de Feira de Santana, Feira de Santana, Brazil
\item[$^{24}$] Universidade Federal de Campina Grande, Centro de Ciencias e Tecnologia, Campina Grande, Brazil
\item[$^{25}$] Universidade Federal do ABC, Santo Andr\'e, SP, Brazil
\item[$^{26}$] Universidade Federal do Paran\'a, Setor Palotina, Palotina, Brazil
\item[$^{27}$] Universidade Federal do Rio de Janeiro, Instituto de F\'\i{}sica, Rio de Janeiro, RJ, Brazil
\item[$^{28}$] Universidad de Medell\'\i{}n, Medell\'\i{}n, Colombia
\item[$^{29}$] Universidad Industrial de Santander, Bucaramanga, Colombia
\item[$^{30}$] Charles University, Faculty of Mathematics and Physics, Institute of Particle and Nuclear Physics, Prague, Czech Republic
\item[$^{31}$] Institute of Physics of the Czech Academy of Sciences, Prague, Czech Republic
\item[$^{32}$] Palacky University, Olomouc, Czech Republic
\item[$^{33}$] CNRS/IN2P3, IJCLab, Universit\'e Paris-Saclay, Orsay, France
\item[$^{34}$] Laboratoire de Physique Nucl\'eaire et de Hautes Energies (LPNHE), Sorbonne Universit\'e, Universit\'e de Paris, CNRS-IN2P3, Paris, France
\item[$^{35}$] Universit\'e Grenoble Alpes, CNRS, Grenoble Institute of Engineering Universit\'e Grenoble Alpes, LPSC-IN2P3, 38000 Grenoble, France
\item[$^{36}$] Universit\'e Paris-Saclay, CNRS/IN2P3, IJCLab, Orsay, France
\item[$^{37}$] Bergische Universit\"at Wuppertal, Department of Physics, Wuppertal, Germany
\item[$^{38}$] Karlsruhe Institute of Technology (KIT), Institute for Experimental Particle Physics, Karlsruhe, Germany
\item[$^{39}$] Karlsruhe Institute of Technology (KIT), Institut f\"ur Prozessdatenverarbeitung und Elektronik, Karlsruhe, Germany
\item[$^{40}$] Karlsruhe Institute of Technology (KIT), Institute for Astroparticle Physics, Karlsruhe, Germany
\item[$^{41}$] RWTH Aachen University, III.\ Physikalisches Institut A, Aachen, Germany
\item[$^{42}$] Universit\"at Hamburg, II.\ Institut f\"ur Theoretische Physik, Hamburg, Germany
\item[$^{43}$] Universit\"at Siegen, Department Physik -- Experimentelle Teilchenphysik, Siegen, Germany
\item[$^{44}$] Gran Sasso Science Institute, L'Aquila, Italy
\item[$^{45}$] INFN Laboratori Nazionali del Gran Sasso, Assergi (L'Aquila), Italy
\item[$^{46}$] INFN, Sezione di Catania, Catania, Italy
\item[$^{47}$] INFN, Sezione di Lecce, Lecce, Italy
\item[$^{48}$] INFN, Sezione di Milano, Milano, Italy
\item[$^{49}$] INFN, Sezione di Napoli, Napoli, Italy
\item[$^{50}$] INFN, Sezione di Roma ``Tor Vergata'', Roma, Italy
\item[$^{51}$] INFN, Sezione di Torino, Torino, Italy
\item[$^{52}$] Istituto di Astrofisica Spaziale e Fisica Cosmica di Palermo (INAF), Palermo, Italy
\item[$^{53}$] Osservatorio Astrofisico di Torino (INAF), Torino, Italy
\item[$^{54}$] Politecnico di Milano, Dipartimento di Scienze e Tecnologie Aerospaziali , Milano, Italy
\item[$^{55}$] Universit\`a del Salento, Dipartimento di Matematica e Fisica ``E.\ De Giorgi'', Lecce, Italy
\item[$^{56}$] Universit\`a dell'Aquila, Dipartimento di Scienze Fisiche e Chimiche, L'Aquila, Italy
\item[$^{57}$] Universit\`a di Catania, Dipartimento di Fisica e Astronomia ``Ettore Majorana``, Catania, Italy
\item[$^{58}$] Universit\`a di Milano, Dipartimento di Fisica, Milano, Italy
\item[$^{59}$] Universit\`a di Napoli ``Federico II'', Dipartimento di Fisica ``Ettore Pancini'', Napoli, Italy
\item[$^{60}$] Universit\`a di Palermo, Dipartimento di Fisica e Chimica ''E.\ Segr\`e'', Palermo, Italy
\item[$^{61}$] Universit\`a di Roma ``Tor Vergata'', Dipartimento di Fisica, Roma, Italy
\item[$^{62}$] Universit\`a Torino, Dipartimento di Fisica, Torino, Italy
\item[$^{63}$] Benem\'erita Universidad Aut\'onoma de Puebla, Puebla, M\'exico
\item[$^{64}$] Unidad Profesional Interdisciplinaria en Ingenier\'\i{}a y Tecnolog\'\i{}as Avanzadas del Instituto Polit\'ecnico Nacional (UPIITA-IPN), M\'exico, D.F., M\'exico
\item[$^{65}$] Universidad Aut\'onoma de Chiapas, Tuxtla Guti\'errez, Chiapas, M\'exico
\item[$^{66}$] Universidad Michoacana de San Nicol\'as de Hidalgo, Morelia, Michoac\'an, M\'exico
\item[$^{67}$] Universidad Nacional Aut\'onoma de M\'exico, M\'exico, D.F., M\'exico
\item[$^{68}$] Universidad Nacional de San Agustin de Arequipa, Facultad de Ciencias Naturales y Formales, Arequipa, Peru
\item[$^{69}$] Institute of Nuclear Physics PAN, Krakow, Poland
\item[$^{70}$] University of \L{}\'od\'z, Faculty of High-Energy Astrophysics,\L{}\'od\'z, Poland
\item[$^{71}$] Laborat\'orio de Instrumenta\c{c}\~ao e F\'\i{}sica Experimental de Part\'\i{}culas -- LIP and Instituto Superior T\'ecnico -- IST, Universidade de Lisboa -- UL, Lisboa, Portugal
\item[$^{72}$] ``Horia Hulubei'' National Institute for Physics and Nuclear Engineering, Bucharest-Magurele, Romania
\item[$^{73}$] Institute of Space Science, Bucharest-Magurele, Romania
\item[$^{74}$] Center for Astrophysics and Cosmology (CAC), University of Nova Gorica, Nova Gorica, Slovenia
\item[$^{75}$] Experimental Particle Physics Department, J.\ Stefan Institute, Ljubljana, Slovenia
\item[$^{76}$] Universidad de Granada and C.A.F.P.E., Granada, Spain
\item[$^{77}$] Instituto Galego de F\'\i{}sica de Altas Enerx\'\i{}as (IGFAE), Universidade de Santiago de Compostela, Santiago de Compostela, Spain
\item[$^{78}$] IMAPP, Radboud University Nijmegen, Nijmegen, The Netherlands
\item[$^{79}$] Nationaal Instituut voor Kernfysica en Hoge Energie Fysica (NIKHEF), Science Park, Amsterdam, The Netherlands
\item[$^{80}$] Stichting Astronomisch Onderzoek in Nederland (ASTRON), Dwingeloo, The Netherlands
\item[$^{81}$] Universiteit van Amsterdam, Faculty of Science, Amsterdam, The Netherlands
\item[$^{82}$] Case Western Reserve University, Cleveland, OH, USA
\item[$^{83}$] Colorado School of Mines, Golden, CO, USA
\item[$^{84}$] Department of Physics and Astronomy, Lehman College, City University of New York, Bronx, NY, USA
\item[$^{85}$] Michigan Technological University, Houghton, MI, USA
\item[$^{86}$] New York University, New York, NY, USA
\item[$^{87}$] University of Chicago, Enrico Fermi Institute, Chicago, IL, USA
\item[$^{88}$] University of Delaware, Department of Physics and Astronomy, Bartol Research Institute, Newark, DE, USA
\item[$^{89}$] University of Wisconsin-Madison, Department of Physics and WIPAC, Madison, WI, USA
\item[] -----
\item[$^{a}$] Max-Planck-Institut f\"ur Radioastronomie, Bonn, Germany
\item[$^{b}$] also at Kapteyn Institute, University of Groningen, Groningen, The Netherlands
\item[$^{c}$] School of Physics and Astronomy, University of Leeds, Leeds, United Kingdom
\item[$^{d}$] Fermi National Accelerator Laboratory, Fermilab, Batavia, IL, USA
\item[$^{e}$] Pennsylvania State University, University Park, PA, USA
\item[$^{f}$] Colorado State University, Fort Collins, CO, USA
\item[$^{g}$] Louisiana State University, Baton Rouge, LA, USA
\item[$^{h}$] now at Agenzia Spaziale Italiana (ASI).\ Via del Politecnico 00133, Roma, Italy
\item[$^{i}$] now at ECAP, FAU Erlangen-N\"urnberg, Erlangen, Germany
\item[$^{j}$] now at Graduate School of Science, Osaka Metropolitan University, Osaka, Japan
\item[$^{k}$] Institut universitaire de France (IUF), Paris, France
\end{description}
}

\begin{abstract}
We present measurements of the atmospheric depth of the shower maximum \xmax, inferred for the first time on an event-by-event level using the Surface Detector of the Pierre Auger Observatory.
Using deep learning, we were able to extend measurements of the \xmax distributions up to energies of 100~EeV ($10^{20}$~eV), not yet revealed by current measurements, providing new insights into the mass composition of cosmic rays at extreme energies.
Gaining a 10-fold increase in statistics compared to the Fluorescence Detector data, we find evidence that the rate of change of the average \xmax with the logarithm of energy features three breaks at $6.5\pm0.6~(\mathrm{stat})\pm1~(\mathrm{sys})$~EeV, $11\pm 2~(\mathrm{stat})\pm1~(\mathrm{sys})$~EeV, and $31\pm5~(\mathrm{stat})\pm3~(\mathrm{sys})$~EeV, in the vicinity to the three prominent features (ankle, instep, suppression) of the
cosmic-ray flux.
The energy evolution of the mean and standard deviation of the measured \xmax distributions indicates that the mass composition becomes increasingly heavier and purer, thus being incompatible with a large fraction of light nuclei between 50~EeV and 100~EeV.
\end{abstract}

\maketitle

\hspace{10em}
\clearpage
\section{\label{sec:intro}Introduction}
The arrival directions, energy spectrum, and mass composition are the three important pillars of cosmic ray research.
A sound interpretation of the three measurements and their energy dependence, both individually and jointly, is pivotal for a deep understanding of the nature of cosmic rays, including their origin and propagation, and enables the study of astrophysical models.
With energies larger than $1\,$EeV ($10^{18}\,$eV), ultra-high-energy cosmic rays (UHECRs) are the most energetic particles ever measured by humankind.
One of the lasting puzzles is the origin of the suppression of the cosmic-ray flux observed at around 50~EeV\cite{Hires2008, Auger_PRL_spectrum_2008, TA_energy_spectrum, the_pierre_auger_collaboration_features_2020}.
A precise measurement of the UHECR mass composition can deliver insights into whether the suppression is caused by the interaction of the particles with the cosmic background photons~\cite{greisen_end_1966, zatsepin_upper_1966}, a sign of the maximum energy reached in cosmic accelerators~\cite{Allard_2008}, or a combination of both~\cite{combined_fit_auger, combined_fit_eleonora}.
Due to the low flux at ultra-high energies, the primary composition cannot be measured directly but can only be studied by indirectly analyzing the properties of the induced air showers.
Information on the primary mass can be obtained by measuring the atmospheric depth of the shower maximum \xmax, the depth at which the number of secondary particles reaches its maximum.
Investigating the measured \xmax distribution, as a function of energy, in terms of its mean and standard deviation (fluctuations), \xmaxmu and \xmaxsigma, enables us to study the UHECR mass composition~\cite{abraham_pierre_auger_collaboration_measurement_2010, pierre_auger_collaboration_depth_2014, aab_pierre_auger_collaboration_depth_2014}.
Heavier particles feature, on average, a smaller \xmax since more sub-showers are created sharing the primary energy.
This results in a maximum higher in the atmosphere and motivates the investigation of the first moment \xmaxmu of the distribution.
Further, the shower-to-shower fluctuations, i.e., the second moment \xmaxsigma of the distribution, is also mass-sensitive.
Due to the smaller cross-section and the development of fewer sub-showers, cascades induced by lighter primary particles are subject to larger fluctuations.
The fluctuations \xmaxsigma are sensitive to both the primary mass and the degree of mixing of the primary beam~\cite{interpretation_auger_jcap}, compared to \xmaxmu, almost insensitive to the uncertainties in the hadronic interaction models.

Using fluorescence telescopes, \xmax can be directly reconstructed by observing the longitudinal shower development.
Nevertheless, due to the observations being confined to dark and moonless nights, the duty cycle is limited.
In contrast, sparse surface-detector arrays have a duty cycle close to $100\%$ and sample the secondary shower particles at the ground.
Thus, they cannot directly observe \xmax, making its reconstruction challenging.
However, information about the shower development and \xmax is contained in the lateral number density and distribution of arrival times of particles reaching the ground.
By studying the risetimes of the time-dependent signals, conclusions on the average composition have already been drawn in the past~\cite{aab_pierre_auger_collaboration_inferences_2017}.
However, to infer the UHECR mass composition beyond mere \xmaxmu measurements, more sophisticated methods are needed to fully exploit the complex data.
The advent of deep learning~\cite{lecun2015deep, Goodfellow-et-al-2016} provides new analysis techniques for large and complex data sets.
First approaches have already been successfully applied to LHC data~\cite{cranmer_dl_lhc_review} and physics in general~\cite{dlfpr}.
The recent progress offers supplementary and improved reconstruction algorithms for neutrino~\cite{abbasi_convolutional_2021, km3net_dl, gal_plane_2023} and cosmic-ray observatories~\cite{erdmann_deep_2018}.
This includes the deep-learning-based reconstruction of \xmax~\cite{xmax_wcd, glombitza_icrc_19, glombitza_icrc_21, glombitza_icrc_23} and muon signals~\cite{pao_muon_dnn} using the temporal structure of signals measured by the Surface Detector of the Pierre Auger Observatory.

In this work, the novel reconstruction technique is used for the first time to study the mass composition of UHECRs in terms of \xmaxmu and \xmaxsigma in the energy range from 3 to 100~EeV.
With about $50{,}000$ events, this is the most comprehensive study of the UHECR mass composition and the first measurement of \xmaxsigma beyond 50~EeV.
A comprehensive discussion of the analysis, including the technical details of the analysis, is given in an accompanying publication~\cite{wcd_dnn_prd}.

\section{\label{sec:method}Methodology}
In the past two decades, our understanding of UHECRs has grown enormously due to the construction of the Pierre Auger Observatory~\cite{auger_nim} and the Telescope Array Project~\cite{ta_kawai}.
The Pierre Auger Observatory is the world's largest UHECR experiment and a hybrid instrument combining surface detectors and fluorescence telescopes to measure cosmic-ray-induced air showers.
In total, 1660 water-Cherenkov detectors, spanning 3000~km$^2$, are arranged in a triangular 1500-meter-grid and form the Surface Detector (SD) --- the centerpiece of the Observatory with a duty cycle close to 100\%.
The SD is overlooked by 24 telescopes located at four sites that form the Fluorescence Detector (FD).
Additionally, three high-elevation telescopes overlook an infilled array of 61 stations with 750~m spacing that enable measurements below 3~EeV.
The requirement for dark and moonless nights limits the duty cycle of the FD to about $15\%$.

The typical size of an air-shower footprint with $E>10$~EeV amounts to tens of $\mathrm{km}^2$, and it usually triggers more than ten stations of the SD.
Each station is equipped with three photomultiplier tubes (PMTs) that record the time-dependent responses to shower particles digitized and sampled in steps of $25$~ns.
The resulting three traces are then calibrated in units of VEM (vertical equivalent muons), i.e., the average signal produced by muons traversing the detector vertically, provided by an in-situ calibration on a minute timescale using atmospheric muons.
Several station-level measurements characterize each event in our analysis: the arrival time of the first particles at the respective station and, for each PMT, a trace of $3~\upmu$s time length (120 time steps) containing the signal.

In this work, we use two different data sets:
a hybrid data set, featuring both an FD and SD reconstruction used to calibrate the reconstruction algorithm to the \xmax scale of the FD, and the full SD data set for performing the high-statistics measurement of \xmaxmu and \xmaxsigma.
We only select events with an energy $E_\mathrm{SD}>3$~EeV to ensure full trigger efficiency~\cite{sd_reco}, require a zenith angle $\theta<60\grad$, and a hexagon of working stations around the station with the largest signal~\cite{abraham_pierre_auger_collaboration_trigger_2010}.
Furthermore, a fiducial SD cut is applied~\cite{wcd_dnn_prd} to ensure an unbiased \xmax measurement, accepting only events inside a zenith-angle range where the absolute \xmax reconstruction bias of the DNN is smaller than $10$~\gcm.
After selection, the SD data set comprises $48{,}824$ events collected between 1 January 2004 and 31 August 2018.
For the calibration of the novel algorithm, hybrid events featuring both an FD and SD reconstruction are used.
We accept only FD events with good atmospheric conditions and small uncertainties on the observed shower profile.
In particular, we reject events with \xmax reconstructed outside the telescope field of view.
To avoid a selection bias, as \xmax depends on the primary mass, we apply a fiducial cut that ensures uniform acceptance for most of the \xmax distribution~\cite{aab_pierre_auger_collaboration_depth_2014}.
$1642$ hybrid events remain after selection.

\begin{figure}[t!]
    \centering
    \includegraphics[width=0.44\textwidth]{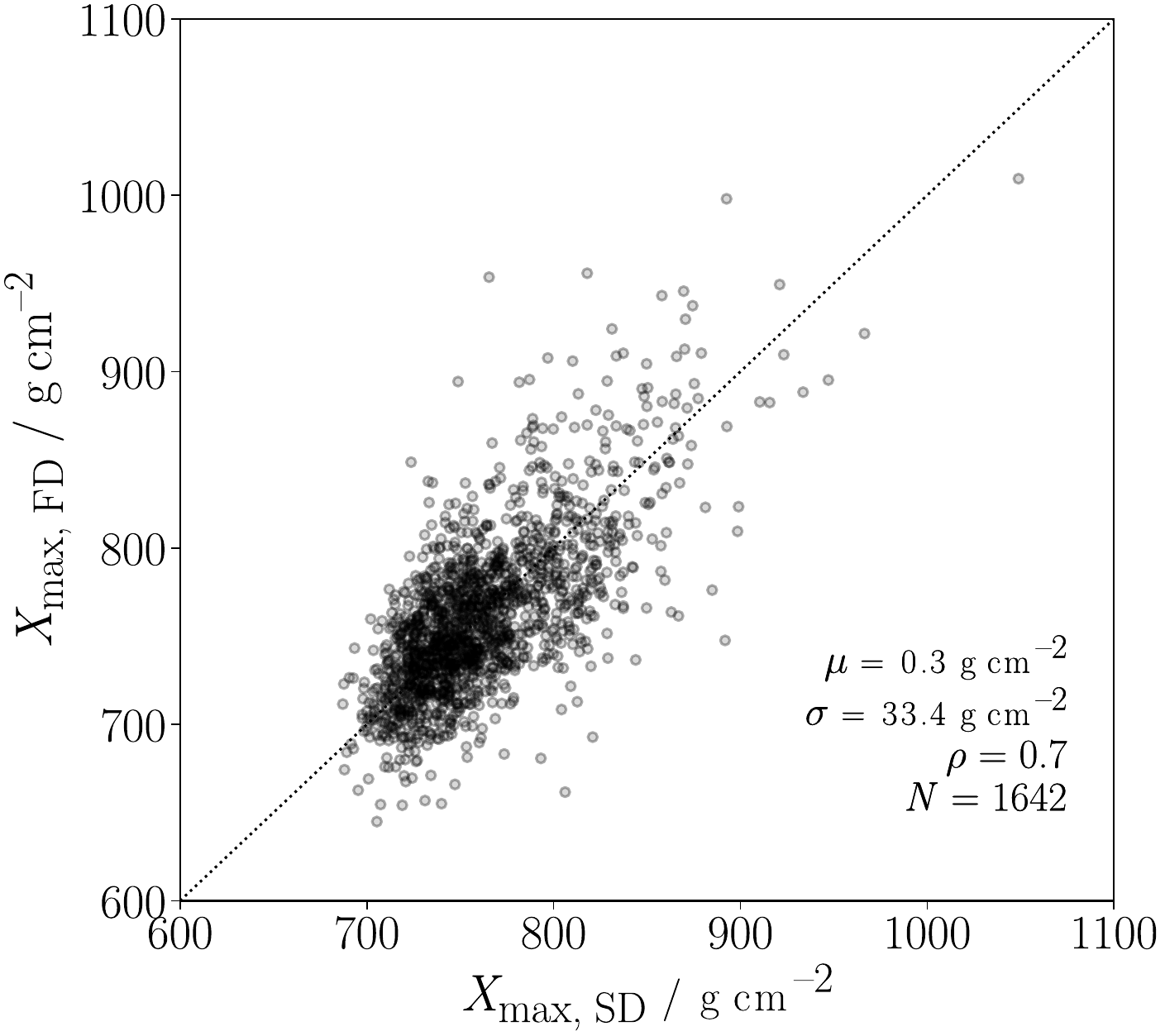}
    \caption{\small{\label{fig:fd_dnn}Application of the DNN to hybrid data. Correlation between fluorescence observations of the FD and DNN predictions using SD data after calibration. The 1642 events show a clear correlation of $\rho = 0.7$ and a bias $\mu<1~\gcm$.}}
\end{figure}

Information on the primary particle mass is encoded in the temporal structure of the recorded SD signals, i.e., the signal traces and arrival times~\cite{aab_pierre_auger_collaboration_inferences_2017, PhysRevD_90_012012}.
The \xmax reconstruction applied in this work is based on a deep neural network (DNN) to exploit the patterns of different shower components in the time-resolved particle density.
For example, muons usually produce signal spikes, while the signals from each electron, positron, and photon are individually smaller and are spread out in time because of multiple scattering~\cite{AVE_univ}.
In the first part, the shape of signal traces is analyzed using long-short term memory (LSTM) layers~\cite{hochreiter_long_1997}, and in the second part, the spatial distribution of the signal footprint induced on the SD grid is exploited using convolutional layers~\cite{hoogeboom_hexaconv_2018}.
The DNN was trained using the simulated detector responses~\cite{mc_library} of $400{,}000$ showers induced by proton, helium, oxygen, and iron with energies from 1 to 160~EeV.
The showers were simulated with CORSIKA~\cite{heck_corsika_1998} using the \epos interaction model.
For more details on the algorithm, we refer to Ref.~\cite{xmax_wcd}.

\begin{figure*}[ht!]
    \begin{centering}
        \begin{subfigure}[b]{0.485\textwidth}
            \begin{centering}
                \includegraphics[width=\textwidth,trim=0.45cm 0.4cm 2.05cm 0.15cm, clip]{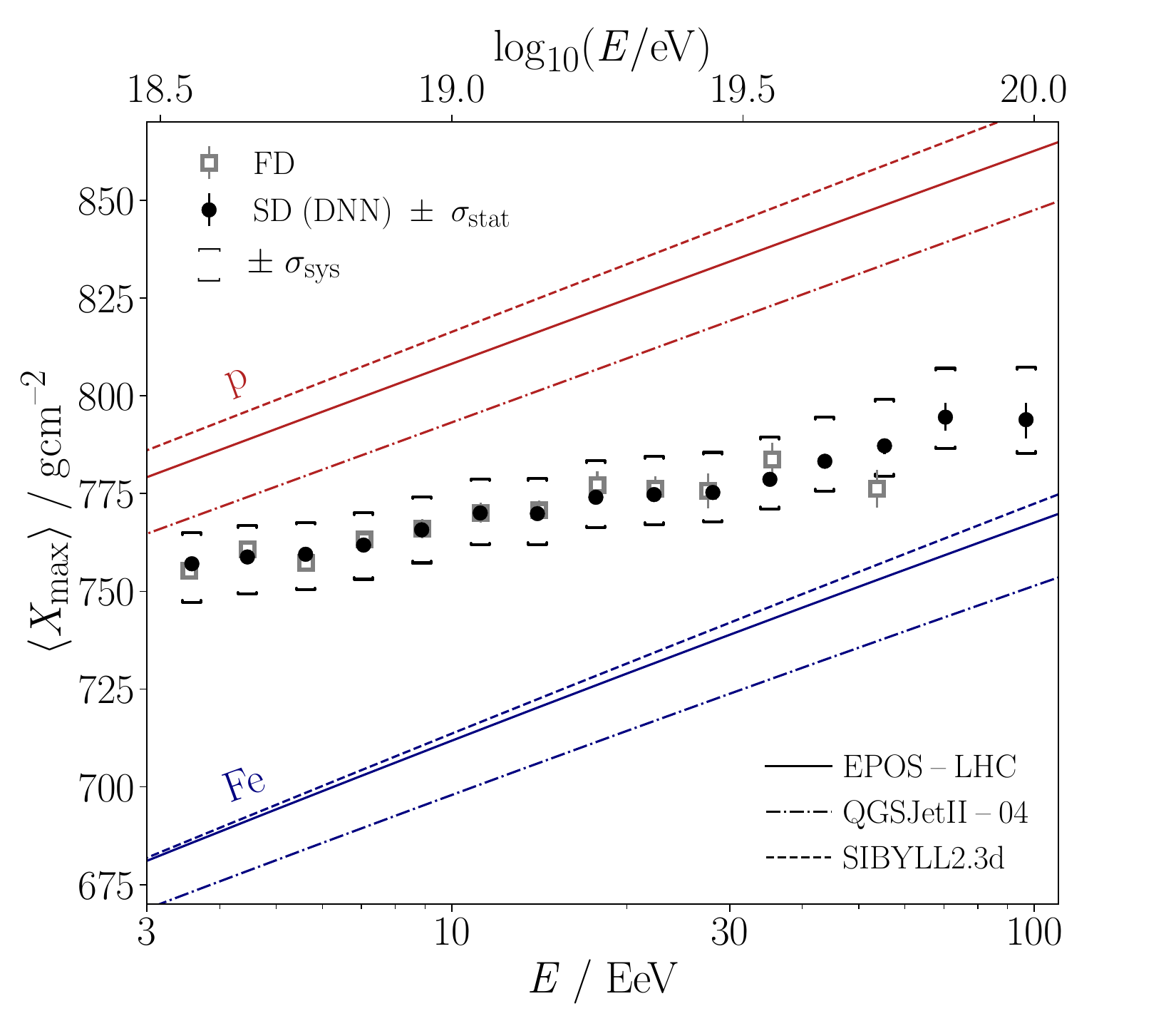}
                \subcaption{}
                \label{fig:1st_moment_dnn_fd}
            \end{centering}
        \end{subfigure}
        \begin{subfigure}[b]{0.485\textwidth}
            \begin{centering}
                \includegraphics[width=\textwidth,trim=0.45cm 0.4cm 2.05cm 0.15cm, clip]{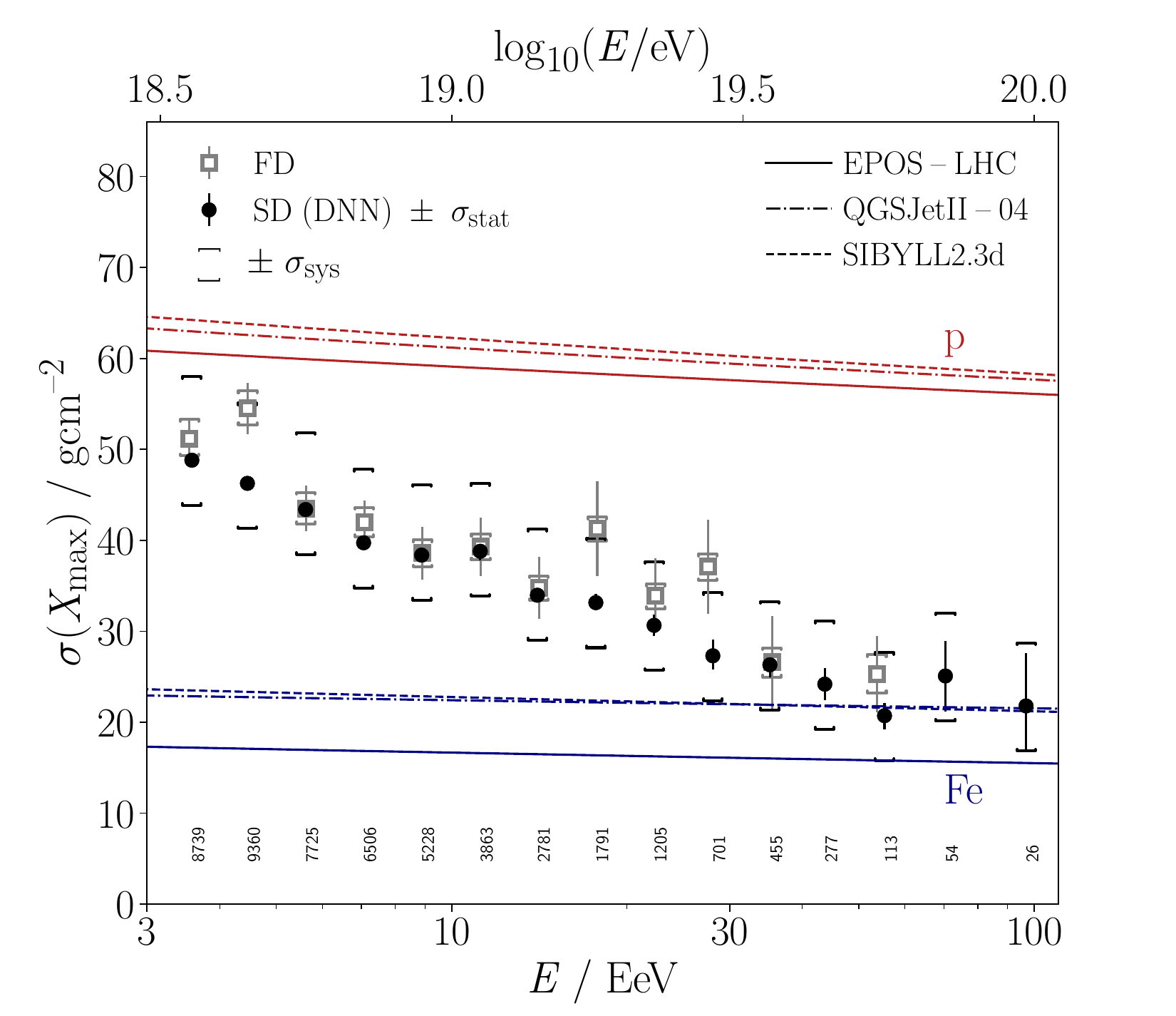}
                \subcaption{}
                \label{fig:2nd_moment_dnn_fd}
            \end{centering}
        \end{subfigure}
    \end{centering}
    \caption{\label{fig:moments_sd_fd}\small{Energy evolution of (a) the mean depth of shower maximum \xmaxmu and (b) the fluctuations of shower maximum \xmaxsigma as determined using the FD reconstruction (grey open squares)~\cite{yushkov_mass_2019} and the SD-based DNN predictions (black circles). Red (blue) lines indicate expectations for a pure proton (iron) composition for various hadronic models. The number of events in each bin is indicated in panel (b).}}
\end{figure*}


After correcting~\cite{wcd_dnn_prd} the reconstruction for aging effects~\cite{PierreAugerlongterm, AbdulHalim:2023iN} and temporal variations, we use hybrid data to calibrate the SD-based DNN algorithm to the FD \xmax scale 
Since UHECRs feature energies above what can be reached with human-built accelerators, air-shower simulations make use of extrapolated collider data and phenomenological modeling that differ for each hadronic interaction model.
As fluorescence telescopes directly observe \xmax, they offer the possibility of removing the dependence of the SD-based algorithm on the particular interaction model and significantly reduce the systematic uncertainties of the \xmaxmu measurement.
By studying the difference between the DNN predictions and the FD observations, we observe an offset of $(-31.7 \pm 0.7)~\gcm$ compatible within uncertainties to be independent of energy ($\Delta \xmax < 6~\gcmd$), as determined by a fit.
The observed offset is larger than the expected differences by up to $-15~\gcm$ from studies using various hadronic interaction models~\cite{xmax_wcd, wcd_dnn_prd}.
This indicates that the current generation of hadronic interaction models may not describe the measured data entirely, which is consistent with previous analyses that suggest inadequacies in the description of muon profiles~\cite{PhysRevD_90_012012, Aab_2016_test_models, the_pierre_auger_collaboration_measurement_muon_fluctuations, aab_pierre_auger_collaboration_inferences_2017}, as well as the longitudinal profiles in general~\cite{jakub_2024testing}.

In \Cref{fig:fd_dnn}, the correlation between the \xmax reconstruction of the DNN and the FD is shown after calibration and is a significant improvement compared to previous analyses~\cite{aab_pierre_auger_collaboration_inferences_2017}.
The found correlation and resolution of the DNN are in excellent agreement with simulation studies~\cite{xmax_wcd} verifying the reconstruction and indicating that the fluctuations are well modeled.
This can be expected, as the shower fluctuations are significantly driven by the fluctuations of the first interaction~\cite{auger_cross_section}, which is relatively similar across hadronic interaction models, and additionally, the relative fluctuations of the number of muons seem to be properly modeled~\cite{the_pierre_auger_collaboration_measurement_muon_fluctuations}.

\section{\label{sec:measurement}Results and discussion}
To investigate the evolution of the UHECR mass composition, we study the first and second moments of the \xmax distributions in \Cref{fig:moments_sd_fd} measured by the SD as a function of energy $E$.
We use bins of $\Delta \log_{10}(E/\mathrm{EeV}) = 0.1$ and an integral bin beyond $10^{19.9}$~eV.
The grey open squares denote FD measurements~\cite{yushkov_mass_2019} of the same data-taking period, and black circles the SD-based DNN measurement of this work, extending the \xmax measurements to the highest energies.
Whereas vertical bars indicate statistical uncertainties obtained via bootstrapping, brackets denote systematic uncertainties.
The red (blue) lines mark predictions~\cite{domenico_reinterpreting_2013} from three hadronic interaction models~\cite{pierog_epos_2015, ostapchenko_qgsjet-ii_2006, sibyll} for a pure proton (iron) composition.
The systematic uncertainties of \xmaxmu range from $9~\gcm$ to $13~\gcm$ and are dominated by the hybrid calibration and the uncertainty of the FD \xmax scale.
The systematic uncertainties of the \xmaxsigma measurement are dominated by the composition bias of the energy measurement and the interaction-model bias of the DNN and are in the order of $\pm6~\gcm$.
This bias was conservatively estimated using a simulation study with various realistic composition scenarios, the measured UHECR energy spectrum~\cite{the_pierre_auger_collaboration_features_2020}, and by considering systematic uncertainties on the reconstruction~\cite{wcd_dnn_prd}.

The \xmaxmu measured with the SD shows excellent agreement with FD observations as shown in \Cref{fig:1st_moment_dnn_fd}.
The measurement shows a transition from a relatively light to a heavier composition, confirming the observation of previous analyses~\cite{abraham_pierre_auger_collaboration_measurement_2010, pierre_auger_collaboration_depth_2014, yushkov_mass_2019, aab_pierre_auger_collaboration_inferences_2017, fd_icrc_23} and extending our measurements to 100~EeV.
As shown in \Cref{fig:2nd_moment_dnn_fd}, with rising energy, the fluctuations diminish and agree well with previous FD measurements.
The observation of decreasing \xmaxsigma implies that besides becoming heavier, the mass composition also has to be rather pure.
This yields a consistent interpretation~\cite{wcd_dnn_prd} of the primary UHECR composition when combined with measurements of \xmaxmu.
The small fluctuations disfavor a substantial fraction of light particles at the highest energies and, at the same time, indicate that the observed suppression in the energy spectrum cannot be entirely ascribed to effects of extragalactic propagation~\cite{combined_fit_auger, combined_fit_eleonora}.

\begin{figure}[t!]
    \centering
    \includegraphics[width=0.49\textwidth,trim=0.85cm 0.8cm 0.9cm 0.8cm, clip]{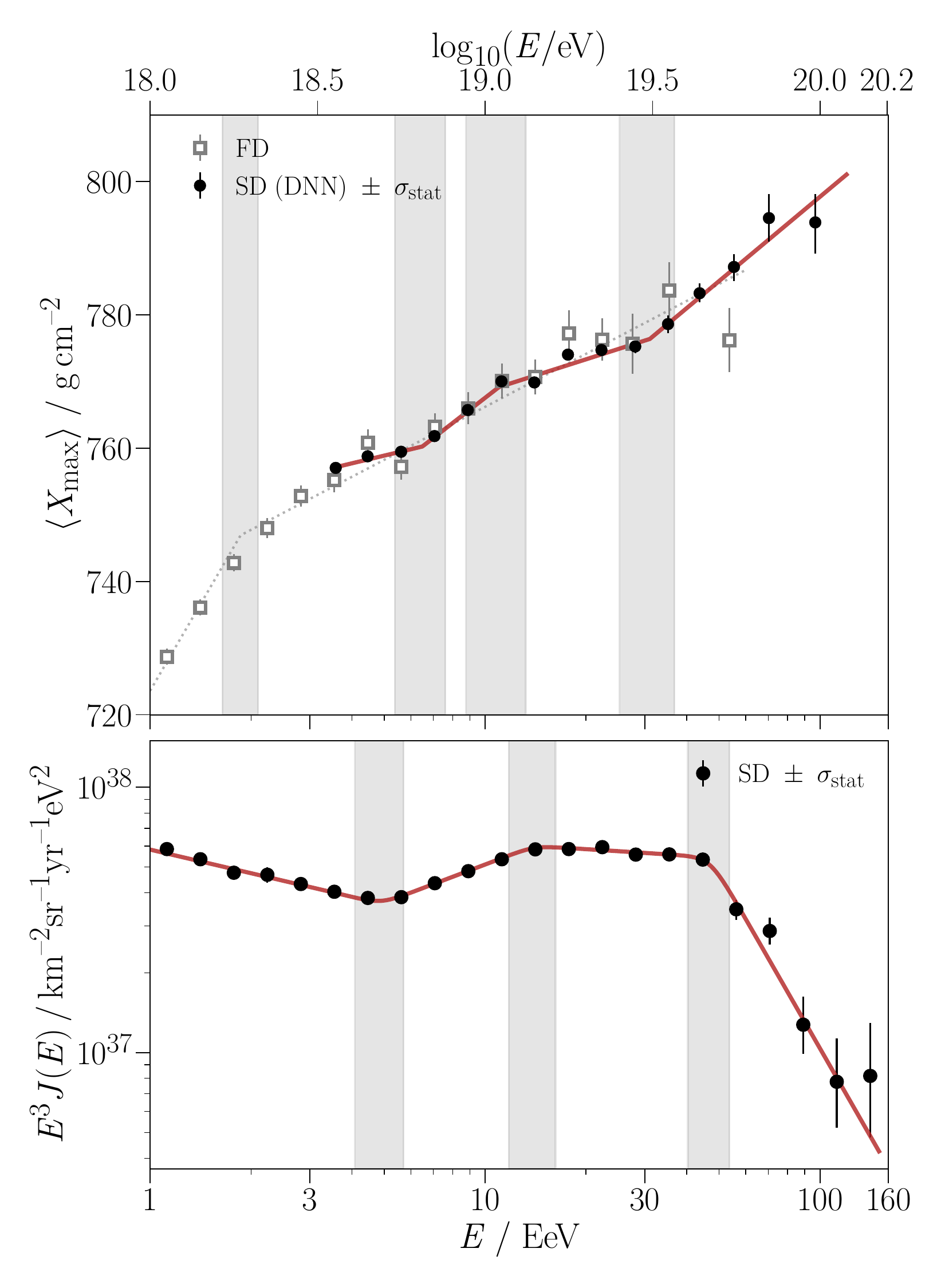}
    \caption{\small{
    Positions of breaks in the elongation rate compared to energy spectrum features. Top: Evolution of \xmaxmu as a function of energy for the SD (black) and the FD (grey)~\cite{yushkov_mass_2019}. The red line indicates the elongation model found using the SD, and the dotted grey line using the FD. Bottom: Combined energy spectrum~\cite{Abreu_2021_spectrum} as measured using the SD 1500~m array and the low energy 750~m infill array of the Observatory.
    Grey regions indicate the uncertainties in the energy of the found breaks and features.
    }}
    \label{fig:elong_rate_spectrum}
\end{figure}

A change in the composition of the primary mass can be studied by investigating the \emph{elongation rate}:
\begin{equation*}
D_{10} \;\hat{=}\; \frac{\mathrm{d}\xmaxmu}{\mathrm{d}\log_{10} E} = \hat{D}_{10} \left( 1 - \frac{\mathrm{d}\langle \ln A\rangle}{\mathrm{d} \ln E } \right ),
\end{equation*}
defined by the change of \xmaxmu in one decade of energy and comparing it to an expected elongation rate $\hat{D}_{10}$ obtained using simulations, which is rather universal across primary masses $A$ and interaction models and ranges from 55 to 60~$\gcmd$.
A linear fit with a constant elongation rate yields $D_{10}=24.1\pm1.2~\gcmd$, in good agreement with the FD measurements in this energy range $\left((26\pm2)~\gcmd\right)$, but does not describe well our data with $\chi^2/\mathrm{dof} = 46.7 / 13$.
Due to the significant increase in statistics, we find evidence for a distinctive structure in the elongation rate when studying the evolution of \xmaxmu using functions piece-wise linear in $\log(E/\mathrm{eV})$.
The observed elongation rate model, shown as a red line in the top panel of \Cref{fig:elong_rate_spectrum}, features three breaks ($\chi^2/ \mathrm{dof} = 10.4/7$).
Using Wilks' theorem, we compared this model with the null hypothesis of a constant elongation rate and found that we can reject the constant elongation rate model at a statistical significance of $4.6\sigma$.
Considering energy-dependent systematic uncertainties, the significance level for rejecting a constant elongation rate reduces to $4.4\sigma$.
We furthermore studied the compatibility of the FD data with our new elongation rate model and observed a good agreement ($\chi^2 / \mathrm{dof} = 12.8 / 12$).\\
The null hypothesis of a model describing the SD with only two breaks at lower energies ($E_1, E_2$), positioned close to the ankle and instep, can be rejected at a statistical significance level of $3.3\sigma$ using the found elongation rate model and shows a stronger dependence on systematic uncertainties.
A single-break model can be rejected with a significance of $4.4\sigma$ and consistently remains above the $3\sigma$ level when including systematics.
The fitted parameters of the model with three breaks are summarized in \Cref{tab:breaks} together with the positions of the energy of spectrum features measured using the SD and the infill array with 750~m spacing.
As shown as a continuous red line in the top panel of \Cref{fig:elong_rate_spectrum}, the found breaks in the evolution of \xmaxmu are observed close to the ankle, instep, and suppression features of the energy spectrum~\cite{Abreu_2021_spectrum}, shown in the bottom panel of \Cref{fig:elong_rate_spectrum}.
The hatched grey regions denote statistical and systematic uncertainties of the position of the features.
Note that distinct features do not have to emerge at similar energies for an astrophysical interpretation of the energy spectrum and its composition.
For example, the break in the elongation rate observed using the FD of the Observatory around 2~EeV~\cite{pierre_auger_collaboration_depth_2014}, shown as a dotted grey line in the top panel of \Cref{fig:elong_rate_spectrum}, is physically interpreted~\cite{Unger_2015, combined_fit_auger, combined_fit_eleonora} in association with the ankle, which has been discovered at 5~EeV.

Interestingly, the composition model discussed in Ref.~\cite{combined_fit_eleonora} (Fig.~3 and Fig.~6), derived by taking into account astrophysical scenarios, including extragalactic propagation and fitting the energy spectrum measured by the SD and the \xmax distribution observed by the FD, predicts three breaks at positions matching our results.
An investigation of this finding to obtain a detailed understanding of the astrophysical origin of these breaks is ongoing.

We also studied the evolution of \xmaxsigma (see \Cref{fig:2nd_moment_dnn_fd}) to identify a potentially similar underlying structure.
We observe a decrease in fluctuations, while the elongation rate implies a change towards a heavier composition.
Consistently, we find no substantial change in the fluctuations \xmaxsigma at the regions --- between the ankle and the instep, and above the suppression --- where the elongation rate of \xmaxmu is closer to that of a constant composition.
While being compatible with the data ($\chi/\mathrm{ndf} = 10.3/10$), a model featuring three breaks at positions fixed to those found in the elongation rate is statistically not significant ($2.2\sigma$) if compared to a linear decrease in \xmaxsigma.
Note that changes in the primary mass composition are not reflected in the same way in the energy evolution of \xmaxmu and \xmaxsigma~\cite{interpretation_auger_jcap}.
A simple transition between two primary species at a constant rate corresponds to a linear dependence of \xmaxmu on $\log(E)$ but to a non-linear behavior of \xmaxsigma, for which, thus, the application of a broken-line model is inappropriate.
For a larger number of primary species with unknown proportions, a specific model for the interpretation of \xmaxsigma cannot be defined.
More sophisticated studies on the astrophysical origin of the features in the mass composition and the energy spectrum are ongoing and will, jointly with the upgrade of the observatory AugerPrime~\cite{castellina_augerprime_2019}, offer new insights into the nature of UHECRs.

\section{\label{sec_summary}Summary}
We have performed a measurement of \xmaxmu and \xmaxsigma for cosmic rays with energies between 3 and 100~EeV to investigate their mass composition.
The method relies on the time-dependent signals recorded by the SD of the Pierre Auger Observatory.
After training our deep learning model on simulated SD data, we used measured hybrid data to crosscheck and cross-calibrate our algorithm using the FD of the Observatory to remove mismatches between simulations and measured data.
With the calibrated DNN, we obtained a 10-fold increase in the size of the \xmax data set for $E>5$~EeV compared to the FD measurements and found a consistent picture of the \xmaxmu and \xmaxsigma measurements.
At lower energies, our measurements are in excellent agreement with fluorescence observations, indicating a light and mixed mass composition.
At the highest, so far inaccessible energies, we report a purer and heavier composition, confirming the trend suggested by the FD data.
The observation of small fluctuations in \xmax beyond 50~EeV excludes a significant fraction of light nuclei at the highest energies and further excludes the flux suppression to be generated by protons interacting with the cosmic microwave background.
However, this observation is insufficient to disentangle whether the suppression arises from the maximum injection energy at the sources, propagation effects, or a combination of both.
Due to the substantial rise in statistics, we have found evidence at a level of $4.4\sigma$ for a characteristic structure in the evolution of the mass composition beyond a constant elongation rate.
The model describing our data best features three breaks.
Interestingly, the identified breaks in the elongation rate model are observed to be in proximity of the ankle, instep, and suppression features in the energy spectrum, where changes in the spectral index have been reported~\cite{the_pierre_auger_collaboration_features_2020}.
More statistics are needed to study the nature of the identified breaks and, particularly, investigate the existence of the third break.
We have demonstrated the large potential of applying deep neural networks to astroparticle physics, particularly in the analysis of low-level data.
Our approach comprises detailed systematic uncertainties, including the cross-calibration with a complementary detector, highlighting the importance of an independent data set for calibration and validation of these powerful algorithms.

The Pierre Auger Observatory is now being upgraded, which includes the deployment of scintillators and radio antennas on top of each SD station.
The new detectors, combined with the emerging capabilities of machine-learning-based algorithms, offer unique prospects for accurate composition studies~\cite{DNN_prime_niklas, DNN_prime_steffen} and increase our understanding of cosmic rays at ultra-high energies.

\section*{Acknowledgments}

\begin{sloppypar}
The successful installation, commissioning, and operation of the Pierre
Auger Observatory would not have been possible without the strong
commitment and effort from the technical and administrative staff in
Malarg\"ue. We are very grateful to the following agencies and
organizations for financial support:
\end{sloppypar}

\begin{sloppypar}
Argentina -- Comisi\'on Nacional de Energ\'\i{}a At\'omica; Agencia Nacional de
Promoci\'on Cient\'\i{}fica y Tecnol\'ogica (ANPCyT); Consejo Nacional de
Investigaciones Cient\'\i{}ficas y T\'ecnicas (CONICET); Gobierno de la
Provincia de Mendoza; Municipalidad de Malarg\"ue; NDM Holdings and Valle
Las Le\~nas; in gratitude for their continuing cooperation over land
access; Australia -- the Australian Research Council; Belgium -- Fonds
de la Recherche Scientifique (FNRS); Research Foundation Flanders (FWO),
Marie Curie Action of the European Union Grant No.~101107047; Brazil --
Conselho Nacional de Desenvolvimento Cient\'\i{}fico e Tecnol\'ogico (CNPq);
Financiadora de Estudos e Projetos (FINEP); Funda\c{c}\~ao de Amparo \`a
Pesquisa do Estado de Rio de Janeiro (FAPERJ); S\~ao Paulo Research
Foundation (FAPESP) Grants No.~2019/10151-2, No.~2010/07359-6 and
No.~1999/05404-3; Minist\'erio da Ci\^encia, Tecnologia, Inova\c{c}\~oes e
Comunica\c{c}\~oes (MCTIC); Czech Republic -- GACR 24-13049S, CAS LQ100102401,
MEYS LM2023032, CZ.02.1.01/0.0/0.0/16{\textunderscore}013/0001402,
CZ.02.1.01/0.0/0.0/18{\textunderscore}046/0016010 and
CZ.02.1.01/0.0/0.0/17{\textunderscore}049/0008422 and CZ.02.01.01/00/22{\textunderscore}008/0004632;
France -- Centre de Calcul IN2P3/CNRS; Centre National de la Recherche
Scientifique (CNRS); Conseil R\'egional Ile-de-France; D\'epartement
Physique Nucl\'eaire et Corpusculaire (PNC-IN2P3/CNRS); D\'epartement
Sciences de l'Univers (SDU-INSU/CNRS); Institut Lagrange de Paris (ILP)
Grant No.~LABEX ANR-10-LABX-63 within the Investissements d'Avenir
Programme Grant No.~ANR-11-IDEX-0004-02; Germany -- Bundesministerium
f\"ur Bildung und Forschung (BMBF); Deutsche Forschungsgemeinschaft (DFG);
Finanzministerium Baden-W\"urttemberg; Helmholtz Alliance for
Astroparticle Physics (HAP); Helmholtz-Gemeinschaft Deutscher
Forschungszentren (HGF); Ministerium f\"ur Kultur und Wissenschaft des
Landes Nordrhein-Westfalen; Ministerium f\"ur Wissenschaft, Forschung und
Kunst des Landes Baden-W\"urttemberg; Italy -- Istituto Nazionale di
Fisica Nucleare (INFN); Istituto Nazionale di Astrofisica (INAF);
Ministero dell'Universit\`a e della Ricerca (MUR); CETEMPS Center of
Excellence; Ministero degli Affari Esteri (MAE), ICSC Centro Nazionale
di Ricerca in High Performance Computing, Big Data and Quantum
Computing, funded by European Union NextGenerationEU, reference code
CN{\textunderscore}00000013; M\'exico -- Consejo Nacional de Ciencia y Tecnolog\'\i{}a
(CONACYT) No.~167733; Universidad Nacional Aut\'onoma de M\'exico (UNAM);
PAPIIT DGAPA-UNAM; The Netherlands -- Ministry of Education, Culture and
Science; Netherlands Organisation for Scientific Research (NWO); Dutch
national e-infrastructure with the support of SURF Cooperative; Poland
-- Ministry of Education and Science, grants No.~DIR/WK/2018/11 and
2022/WK/12; National Science Centre, grants No.~2016/22/M/ST9/00198,
2016/23/B/ST9/01635, 2020/39/B/ST9/01398, and 2022/45/B/ST9/02163;
Portugal -- Portuguese national funds and FEDER funds within Programa
Operacional Factores de Competitividade through Funda\c{c}\~ao para a Ci\^encia
e a Tecnologia (COMPETE); Romania -- Ministry of Research, Innovation
and Digitization, CNCS-UEFISCDI, contract no.~30N/2023 under Romanian
National Core Program LAPLAS VII, grant no.~PN 23 21 01 02 and project
number PN-III-P1-1.1-TE-2021-0924/TE57/2022, within PNCDI III; Slovenia
-- Slovenian Research Agency, grants P1-0031, P1-0385, I0-0033, N1-0111;
Spain -- Ministerio de Ciencia e Innovaci\'on/Agencia Estatal de
Investigaci\'on (PID2019-105544GB-I00, PID2022-140510NB-I00 and
RYC2019-027017-I), Xunta de Galicia (CIGUS Network of Research Centers,
Consolidaci\'on 2021 GRC GI-2033, ED431C-2021/22 and ED431F-2022/15),
Junta de Andaluc\'\i{}a (SOMM17/6104/UGR and P18-FR-4314), and the European
Union (Marie Sklodowska-Curie 101065027 and ERDF); USA -- Department of
Energy, Contracts No.~DE-AC02-07CH11359, No.~DE-FR02-04ER41300,
No.~DE-FG02-99ER41107 and No.~DE-SC0011689; National Science Foundation,
Grant No.~0450696; The Grainger Foundation; Marie Curie-IRSES/EPLANET;
European Particle Physics Latin American Network; and UNESCO.
\end{sloppypar}

\begin{table}[h!]
\caption{Best-fit parameters with statistical and systematic uncertainties for the identified model that features three changes in the elongation rate ($D_0, D_1, D_2, D_3$) at energies ($E_1, E_2, E_3$) and an offset $b$ of \xmaxmu at 1~EeV. The positions of the features of the energy spectrum~\cite{Abreu_2021_spectrum} are also given.}
\begin{tabular}{ c c c c c c}
    \hline\hline
     parameter & 3-break model & energy spectrum \\
$\mathrm{val} \pm \sigma_\mathrm{stat} \pm \sigma_\mathrm{sys}$ & $\mathrm{val} \pm \sigma_\mathrm{stat} \pm \sigma_\mathrm{sys}$ & $\mathrm{val} \pm \sigma_\mathrm{stat} \pm \sigma_\mathrm{sys}$  \\
    \hline
    $b$~/~\gcm    & $750.5\pm3\pm13$ &  \\
    $D_0$~/~\gcmd & $12\pm5\pm6$   &  \\
    $E_1$~/~EeV   & $6.5\pm0.6\pm1$& $4.9\pm0.1\pm0.8$  \\
    $D_1$~/~\gcmd & $39\pm 5\pm14$  &  \\
    $E_2$~/~EeV   & $11\pm 2\pm 1$ & $14\pm1\pm2$  \\
    $D_2$~/~\gcmd & $16\pm3\pm6$   &  \\
    $E_3$~/~EeV   & $31\pm5\pm3$   & $47\pm 3\pm 6$  \\
    $D_3$~/~\gcmd & $42\pm9\pm12$   &  \\
    \hline\hline
\end{tabular}
\label{tab:breaks}
\end{table}

\bibliographystyle{apsrev4-2}

 http://www.auger.org
\email{auger\_spokespersons@fnal.gov}
\bibliography{PRL}


\end{document}